# Ellipsometry studies of Si/Ge superlattices with embedded Ge dots

Şeref Kalem, Örjan Arthursson, Peter Werner

*Abstract*— **In this paper, we present an analysis for treating the spectroscopic ellipsometry response of Si/Ge superlattices (SL) with embedded Ge dots. Spectroscopic ellipsometry (SE) measurement at room temperature was used to investigate optical and electronic properties of Si/Ge SL which were grown on silicon (Si) wafers having <111> crystallographic orientation. The results of the SE analysis between 200 nm and 1000 nm indicate that the SL system can effectively be described using interdiffusion/intermixing model by assuming a multicrystalline Si and $Si_{1-x}Ge_x$ intermixing layers. The electronic transitions deduced from analysis reveal Si, Ge and alloying related critical energy points.**

## 1. Introduction

Si/Ge superlattices have attracted a significant interest due to a great potential for applications in advanced micro and nano electronics where CMOS compatible electronic and photonic components can be integrated at lower cost and higher efficiency on Si platforms [1, 3]. They could also find applications in sensing and power generation [4]. The hope for the development of a photonic Si based platform is supported by the possibility of creating a quasi direct band gap Si and Ge light sources. Short period Si/Ge superlattices (SL), quantum dots and nanowires can offer such possibilities through changing properties in electronic band structure. Strain induced modification of the band structure and the size reduction effects can favor direct band gap formation and related direct transitions [5, 6]. Such changes would increase oscillator strength for optical transitions and thus leading to effective band gap engineering possibilities which would favor electronic applications in a variety of fields [7, 8].

A number of groups have investigated optical and electronic properties of Si/Ge superlattices as a function of composition, strain and period [9]. Common feature of the observed properties is the splitting, broadening and relative amplitude changes in electronic transitions [10]. In addition to bulk like transitions ($E_1$ and $E_2$), superlattice related transitions were demonstrated between $E_1$ and $E_2$ peaks [11]. These features are accompanied by shifting of electronic transition energies and splitting particularly at high energy critical points (CP) $E_1$ and $E_2$. Both transitions were found to be decreasing in energy with increasing period. Only the lower energy peak exhibited an amplitude enhancement and a broadening with period indicating the evidence for confinement effects. Contrary to these observations claiming the splitting only in high period number SLs, a later work reported a splitting of the $E_2$ peak in short period superlattices [12]. On the other hand, intermixing and alloying at interfaces could introduce further complications in the interpretation of

S. Kalem
TUBITAK-BILGEM National Research Institute of Electronics and Cryptology, Gebze 41470 Kocaeli, Turkey
e-mail: seref.kalem@tubitak.gov.tr)

Ö. Arthursson
Department of Microtechnology and Nanosciences, Chalmers University of Technology, Göteborg, Sweden

P. Werner
Department of Experimental Physics, Max-Planck-Institute, Halle (Saale), Germany

optical response. Both strained and unstrained Si$_{1-x}$Ge$_x$ alloys exhibit composition dependent down-shifts in CP energy and lineshape [13]. The dielectric function of Ge dot layer which was embedded in SiGe alloy was found not to correspond to an alloy or bulk Ge, but affected by confinement [14]. CP energies for Si/Si$_{1-x}$Ge$_x$ alloys and Si/Si$_{1-x}$Ge$_x$ SLs were reported for a range of compositions [15, 16].

True interpretation of Si/Ge SL structures relies on accurate determination of individual layer parameters such as thickness, composition, roughness, crystal properties. SE analysis can provide valuable information about these quantities which are needed for the treatment of the optical response of Si/Ge SL structures.

## 1. Experimental

Si/Ge SL structures with embedded Ge dots were prepared by molecular beam epitaxy (MBE) on phosphorous doped Si <111> wafers which was followed by a Si buffer layer of 200 nm [1, 2]. They consist of ultrathin Si/Ge multilayers (superlattices) whose thicknesses is ranged from few monolayer (ML, where **1ML=a/3$^{1/2}$** ,a is the lattice constant, **a$_{Si}$**=5.43095 Å, **a$_{Ge}$**=6.4613 Å) to few nm. The superlattice structure studied here is composed of a 39 pairs of 9 ML Si and 2 ML Ge layers. Each structure was terminated by a Si cap layer with a thickness of 5 nm. The details of the MBE growth were described elsewhere [1-3]. Spectroscopic Ellipsometer (SE) analysis was performed using a Woollam Ellipsometer between 0.2 - 1.0μm. Details on the transmission electron microscopy (TEM) studies of these structures revealing embedded Ge dots can be found in earlier reports [17].

## 3. Results and discussion

A typical SEM image is shown in Fig. 1 for a wafer consisting of a Si(9ML)/Ge(3ML) SL of 39 periods with a buffer layer of 200 nm Si grown on phosphorous doped Si(111) substrate with a resistivity of ~10-20 Ohm.cm. The wafer has been treated in acid vapor for 30 seconds revealing the presence of Ge dots of about 250 nm in lateral size. The acid vapor used a mixture of HF:HNO$_3$ (7.5:2.5) which is selectively reactive with Si as compared to Ge. Details of this treatment was reported elsewhere [18]. Note that the surface is not a smooth single crystalline layer but resembles that of a multi-crystalline structure. The Ge dots are randomly distributed on the surface and the interspacing between the dots ranges from 0.4 μm to about 2.0 μm.

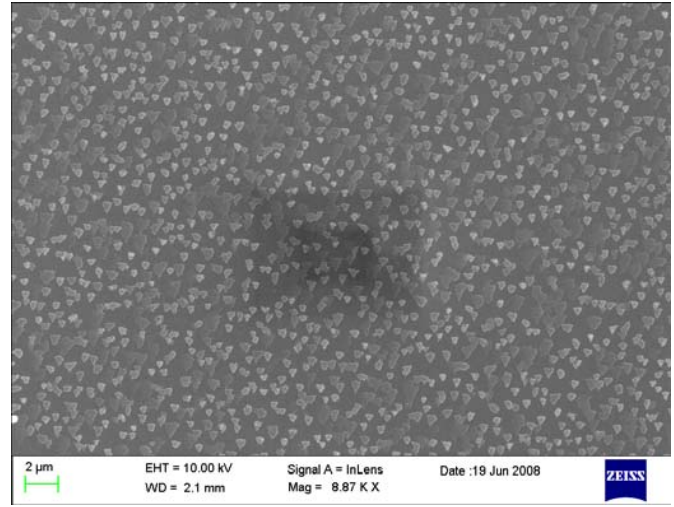

**Fig. 1** SEM image of the surface structure representing a Si(9ML)/Ge(3ML) grown on Si (111) with a buffer layer of Si(200 nm). The surface was exposed for 30 second to vapor of acid mixture consisting of HF:HNO$_3$.

### 3.1 Spectroscopic ellipsometry analysis

Spectroscopic ellipsometry analyses enabled us to determine the critical point energies as well as the information on thickness, crystallinity, roughness and the composition of individual layers, which are particularly valuable for Si/Si$_{1-x}$Ge$_x$ heterostructured multilayer system. Spectroscopic ellipsometry measures the ratio **ρ** of the reflection coefficient **r$_p$** and **r$_s$** (**p** parallel and **s** perpendicular to the plane of incidence, respectively). This can be expressed in terms of the amplitude ratio **tan ψ** and the phase angle **Δ**:

$$\rho = \; (r_p/r_s) = (\tan\psi) \, e^{(i\Delta)} \qquad (1)$$

The complex dielectric function **ε (ω) = ε$_1$(ω) + iε$_2$(ω)** can be derived using the two-phase model [19] where **ε** is expressed by

$$\varepsilon = \sin^2\varphi + \sin^2\varphi \tan^2\varphi \, (1-\rho)^2/(1+\rho)^2$$

where $\varphi$ is the angle of incidence. In Figure 2, the results of SE measurements and analysis are displayed for an as-grown superlattice structure consisting of 9 ML (2.82nm) of Si / 2 ML(0.66nm) of Ge with a period number of 39 and a Si cap layer of 5 nm. The whole structure was grown on Si (111) wafer with a buffer layer of Si having a thickness of 200 nm. A very good fit can be obtained between the experimental data and a generated one using the interdiffusion model [21] as shown in Figures 2(a) and (b) for the cases of two angle of incidence (65° and 75°) for $\psi$ and $\Delta$ between 200 nm and 1000 nm.

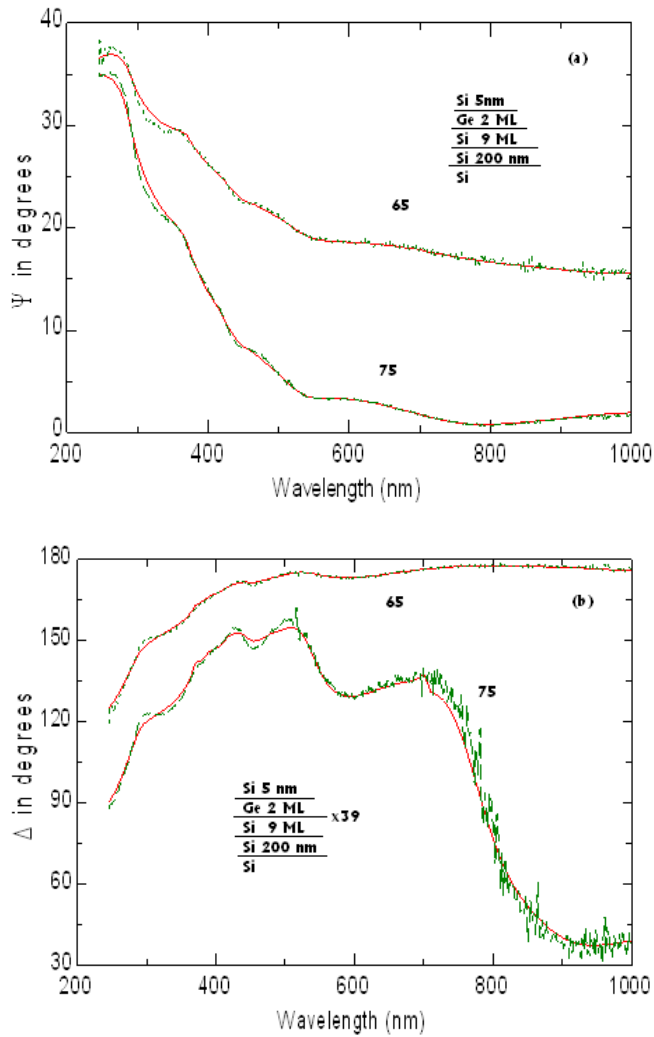

**Fig. 2** Raw spectroscopic ellipsometry data for two angles of incidence 65° and 75°: (a) the amplitude ratio $\psi$ ; (b) the phase angle $\Delta$ and generated model calculation (solid lines) for a Si/Ge superlattice structure consisting of 9ML Si/2ML Ge with 39 periods and a Si cap layer of 5 nm as inserted.

The best fit to experimental data was obtained assuming a polycrystalline Si buffer and Si layers as displayed in Table-1 where the 2nd and 3rd layers repeated 39 times forming a superlattice structure. Actually, TEM investigation revealed the formation of Ge islands or dots consisting of elongated Ge dots with lateral dimensions of more than 100 nm in the case of plane SL structure [2]. The density of these islands in SL structure can be as high as $4 \times 10^{11}$ cm$^{-2}$ [20]. Also, note that SL sample surface consists of an uneven crystalline structure with Ge dots distributed randomly on the surface having interspacing ranging from 0.4 µm to 2.0 µm as shown in Fig. 1. Thus, the surface can be assumed as a multicrystalline like structure consisting of a Si$_{1-x}$Ge$_x$ layer with embedded Ge dots. In line with these considerations, the best fitting requires the presence of a polycrystalline Si superlattice layer of Si$_{1-x}$Ge$_x$ (x=0.036) and a Ge layer of Si$_{1-x}$Ge$_x$ (x=0.5) as a result of intermixing between the adjacent Si and Ge layers.

| TABLE I | | | |
|---|---|---|---|
| **PARAMETERS USED FOR FITTING ELLIPSOMETRY DATA** | | | |
| native oxide | Oxide layer | SiO$_2$ | |
| *cap layer* | 5 nm of Si | Poly SiGe x=0.122  11.470 nm | |
| ***Superlattice*** | **2 ML Ge** | **SiGe  x=0.5  0.093 nm** | **39 periods** |
| | **9 ML Si** | **Poly SiGe x=0.036  2.947 nm** | |
| *Buffer layer* | Si 200 nm | Poly Si  187.406 nm | |
| Wafer | Si wafer | Si wafer  1 mm | |

Although the Si layer thickness is very close to what was predicted, Ge sublayer is thinner which is probably due to interdiffusion effect. The model assumes that a cap layer of polycrystalline Si$_{1-x}$Ge$_x$ (x=0.122) was followed by a SiO$_2$ cap layer of 2.5 nm to obtain a good agreement above the E2 peak energy region. Actually, Raman measurements on Si/Ge structures grown under same conditions on Si (111) indicate a composition x of about 12% (x=0.12) as deduced from the Si-Si vibration at 512 cm⁻¹ in $Si_{1-x}Ge_x$ thus supporting the intermixing as shown in Fig. 3 (More comprehensive studies on Raman light scattering in these alloys can be found in refs [22-24]). The peaks of Ge-Ge vibrational mode at 296 cm⁻¹ and Si-Ge mode at 410 cm⁻¹ have relatively strong intensities compared to an equivalent $Si_{1-x}Ge_x$ (x=0.12). Such peaks are usually found to be mainly confined at the interfaces between Ge and Si layers. These vibrational modes can be explained by the concept of zone-folding of the optical phonons [25]. Weak bands observed between bulk Si-Si and Si-Ge modes should be due to localized Si-Si vibrations in the neighborhood of one or more Ge atom [26]. $Si_3$-Si-$Ge_1$ and $Si_1$-Si-$Ge_3$ bonds could be two examples of such bonding configurations.

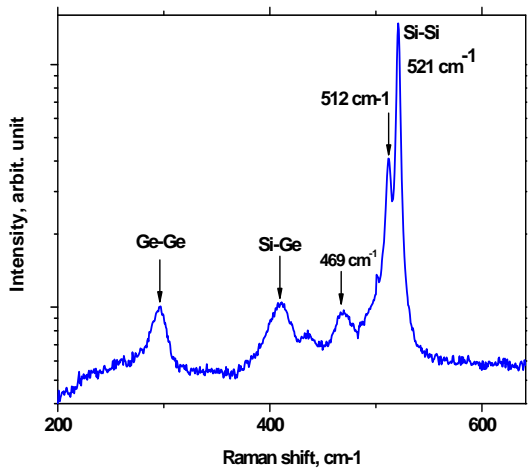

**Fig. 3** Raman light scattering from Si(200nm)/Ge(100nm) heterostructure grown on Si(111) indicating the presence of SiGe alloying. Raman signal was collected using 488 nm excitation line of an Ar+ laser. The arrow at 469 cm⁻¹ indicates the likely position of a Si-Si optical phonon (TO-phonon) of an amorphous like Si matrix [27] [28]. The band at 512 cm-1 is a Si-Si phonon vibration in Si₁₋ₓGeₓ of x=0.12.

### 3.2 Electronic transitions

Second derivative of the dielectric $\varepsilon$ function provides information on the CPs of the electronic transitions as shown in Fig. 4 for a SL structure of a 2 ML Ge/9 ML Si with N=39. The minima in the second derivative spectrum, $d^2\varepsilon(\omega)/d\omega^2$ at 4.35 eV and 3.43 eV would correspond to $E_2$ and $E_1$ transitions of Ge dots and Si layers by analogy to previous work on Si/Ge superlattices [13-15]. Main reason for the attribution is the domination of the E2 CP with increasing Ge total layer thickness. In SLs where the total thickness of Si is much larger than that of Ge, the E1 CP is the dominant feature. Transitions lower than E2 could be referred to as superlattice transitions [29]. These points are actually close to the corresponding transitions of alloys of the same composition, that is x=$t_{Ge}$/($t_{Ge}$+$t_{Si}$), where $t_{Ge}$ and $t_{Si}$ are the thicknesses of Ge and Si sublayers, respectively. The transition at 2.66 eV is close to the values interpolated from bulk Si and Ge. Thus we have good reason to assume that we observe the average transition from SiGe alloy of about x=0.5 [16, 30] originating from interface modes [31]. The CPs at around 3.0 eV could correspond to $E_0^1$ transition in Ge dots

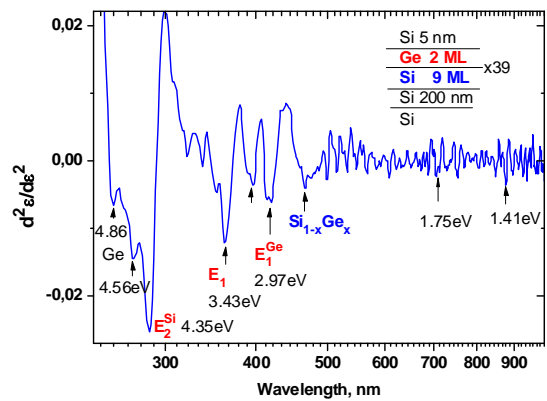

**Fig 4** The second derivative of the dielectric functions of as-grown Si/Ge superlattice with sublayers of Si(9ML)/Ge(2ML) having a period number of N=39. With the increasing number of periodicity, $E_2$ transitions broaden and dominate in the high energy region. There is also a fine triplet structure at around $E_2$.

[14].

We observe that the $E_2$-like transitions split into a doublet with a splitting energy of about 200 meV and the splitting persists in both superlattices contrary to what was observed earlier [32]. The amplitude and the width of this CP was found to be enhanced in the SL with a Ge layer of 3 ML and decreased for a SL of 9 periods. This is a strong evidence for Ge contribution to this peak and particularly to the shoulder at high energy side. The transition at 4.5 eV could be associated with a strained Ge layer as LDA calculations predict [12].

These transitions are located at higher energies as compared to bulk Si and Ge [29,30]. It was recently shown that Ge dots would exhibit large increases in $E_1$ and $E_2$ transition energies [14]. Lateral confinement of carriers due to multicrystalline nature of the Si/Ge SL structure could also be at the origin of such features.

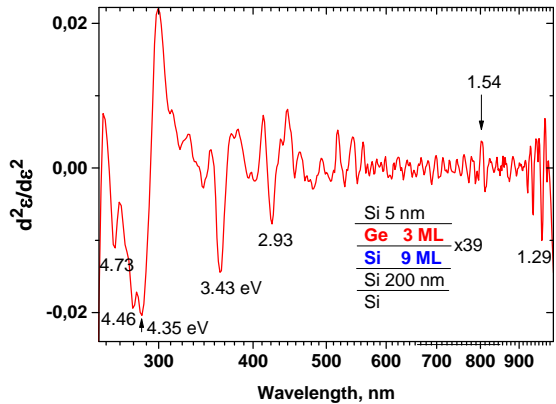

**Fig 5** Second derivative $d^2\varepsilon(\omega)/d\omega^2$ of as-grown Si/Ge superlattice with sublayers of Si(9ML)/Ge(3ML) having a period number of N=39. With the increasing effective layer thickness of Ge, Ge related transitions become stronger at around E1 and E2 transitions. The fine triplet structure is still distinguishable

In Fig. 4 and 5, at lower energies ($<E_1$), the interpretation of the SE spectra is more difficult due to interference effects and noise. In this region, there are weak transitions at around 1.75 eV and 1.41 eV, which can be attributed to unfolded and folded transitions at k=0, respectively [10]. But, when the Ge layer thickness increases from 2 ML to 3 ML, a singularity appears at around 1.3 eV. The photoluminescence observed from the same SL of 9 ML Si/3ML Ge indicates emissions peaks at around 435 nm, 540 nm. These peaks could be related to $E_1$ and SiGe interface modes. The onset of the photoluminescence from about 700 nm to lower energies could be associated with the lower energy transitions observed in SE data in Fig. 5.

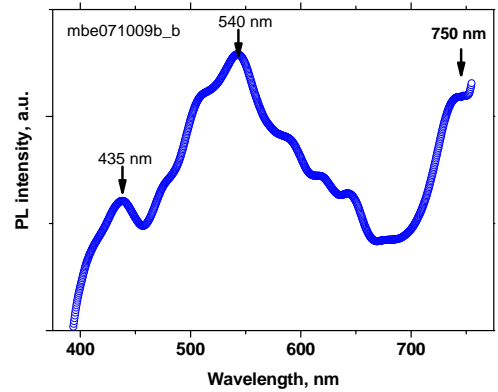

**Figure 6** Photoluminescence of the same SL structure consisting of a Si(9ML)/Ge(3ML) with a period number of N=39.

## 4. Conclusion

Optical response from the Si/Ge superlattice structures having Ge clustering is best treated assuming a multicrystalline and alloying effects. Spectroscopic ellipsometry analysis combined with structural properties is indicative of the presence of alloying (intermixing of Si and Ge superlattice layers) and multicrystalline-like structure due to the presence of Si and Ge rich regions consisting of Ge dots or clusterings. The multicrystalline layers and intermixing of Si and Ge is in accordance with the results of Raman scattering. The amplitude and variations in energetic positions of electronic critical point energies suggest the presence of strong optical absorption onsets indicating the contribution of Ge quantum dots, silicon and $Si_{1-x}Ge_x$ interfacial alloying of Si/Ge superlattice structure to optical transitions. Splitting and broadening of $E_2$ critical point energy is typical for these structures. The results can serve as useful hints in designing and realizing novel photonic devices.

**Acknowledgments** This work was supported by TUBITAK (TBAG) bilateral program under contract No. 107T624, and BMBF German Federal Ministry of Education and Research (Grant No: 03Z2HN12). The authors wish to acknowledge for partially funding of this work to ''MC2 ACCESS'' FP6 EU

Program (Contract No: 026029). We would like to thank A. Frommfeld for supporting the MBE growth, Dr. M. Becker for Raman measurement.